\documentclass[fleqn,twoside]{article}
\usepackage{espcrc2}
\IfFileExists{srcltx.sty}{\usepackage[active]{srcltx}}%
\usepackage{graphicx}
\usepackage[figuresright]{rotating}

\newcommand{\AmS}{{\protect\the\textfont2
  A\kern-.1667em\lower.5ex\hbox{M}\kern-.125emS}}

\title{Sterile dark matter and reionization}

\author{Alexander Kusenko\address{Department of Physics and Astronomy,
University of California, Los Angeles, CA 90095-1547, USA}}

\begin{document}

\begin{abstract}
Sterile neutrinos with masses in the keV range can be the dark matter,
and their emission from a supernova can explain the observed velocities of
pulsars. The sterile neutrino decays could produce the x-ray radiation in
the early universe, which could have an important effect on the formation of
the first stars.  X-rays could ionize gas and could catalyze the production of
molecular hydrogen during the ``dark ages''.  The increased fraction of
molecular hydrogen could facilitate the cooling and collapse of the
primordial gas clouds in which the first stars were formed.  

\vspace{1pc}
\end{abstract}

\maketitle

There are several reasons to expect that cosmological dark matter is made up
of right-handed, or sterile neutrinos.  First, most model of the neutrino
masses postulate the existence of the right-handed fields.  Although it is not
impossible to explain the neutrino masses otherwise, the seesaw
mechanism~\cite{seesaw} offers probably the most natural way of doing it. 
However, the seesaw mechanism can work equally well for the sterile neutrino
masses well above~\cite{seesaw} or well 
below~\cite{deGouvea:2005er,nuMSM,deGouvea:2006gz} the electroweak scale.  Each
of these two possibilities offers a viable scenario for leptogenesis, in which
the lepton asymmetry is generated by either the sterile neutrino
decays~\cite{Fukugita:1986hr}, or by neutrino
oscillations~\cite{baryogenesis}.  The current experimental constraints allow
both possibilities~\cite{sterile_constraints}.  

\begin{figure}[htb]
\includegraphics[width=18pc]{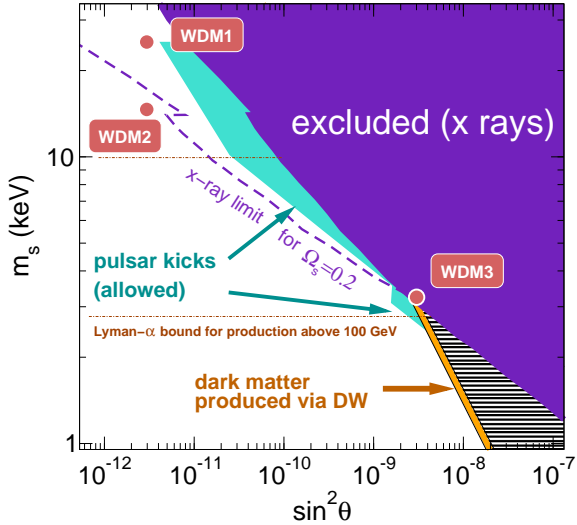}
\caption{The limits on sterile neutrino mass $m$ and mixing angle $\theta$. 
Also shown are the three sets of parameters, WDM1-3, corresponding to the
curves in Fig.~\ref{fig:H2}. The
dashed line shows the x-ray limit if the sterile neutrinos account for the
entire dark matter. If the neutrino oscillations are the only production
mechanism, the corresponding limits are weaker (the solid purple
region)~\cite{Kusenko:2006rh}.  The
Lyman-$\alpha$ lower bound on the dark-matter particle mass depends on the
production mechanism.  This bound is 10~keV~\cite{viel} for the production via
oscillations~\cite{dw}, but it is relaxed considerably if the population of
relic neutrinos originates above the electroweak scale~\cite{Kusenko:2006rh}.  
}
\label{fig:one}
\end{figure}

Second, there are some additional astrophysical hints in favor of the keV
sterile neutrinos.  The observed velocities of pulsars~\cite{Kusenko:review}
can be explained by the emission of sterile neutrinos from a
supernova~\cite{ks97}.
The magnetic field in a hot protoneutron star can grow via the dynamo action
driven by the neutrino cooling at least until the saturation of the linear
regime is achieved, which is for $B\sim 10^{16}$~G~\cite{dt}.  Since the growth
is concentrated in the thin-shell convective zone, the exponentially growing
modes on different sides of the star can develop in an uncorrelated
manner~\cite{Socrates:2004tt}, and the global structure of the field in the
core need not be spherically or axially symmetric during the first seconds of
the supernova.  Hence, the field in the core can have a strong off-centered
dipole component.  The electrons in the hot protoneutron star are polarized by
the magnetic field. Because of this, both the active and the sterile neutrinos
are produced anisotropically, but for the active neutrinos the anisotropy is
washed out by scattering~\cite{eq}. In contrast, the sterile neutrinos escape
without scattering, so their emission asymmetry equals the production
asymmetry, and the recoil can give the neutron star a kick consistent with the
observations~\cite{ks97}.   The range of parameters consistent with this 
explanation of the pulsar kick is shown in Fig.~\ref{fig:one}.  Numerical
calculations of supernova explosion taking into account this mechanism for the
pulsar kick show that the motion of the neutron star causes convection, which
can deposit additional energy beyond the shock, thus helping the supernova
explode~\cite{Fryer}. 
In the early universe sterile neutrinos can be produced from neutrino
oscillations~\cite{dw,Fuller,shi_fuller} or from the inflaton
decay~\cite{shaposhnikov_tkachev} at some sub-GeV temperatures.  They can
also be produced in some other process, for example, in the singlet Majoron
decays at some temperature above the electroweak scale~\cite{Kusenko:2006rh}.
Different production mechanisms generate neutrinos with different momentum
distributions. To be dark matter, the relic population of sterile neutrinos
must be sufficiently cold.  For neutrino oscillations in the absence of lepton
asymmetry, the Lyman-$\alpha $ bound is $m_s>10$~keV~\cite{viel}.  If the
lepton asymmetry of the universe is 0.01 or greater, resonant neutrino
oscillations can generate the dark matter in the form of sterile neutrinos even
for very small mixing angles~\cite{shi_fuller}.  If the sterile neutrinos are
produced above the electroweak scale, their momenta are redshifted, and the
Lyman-$\alpha$ bound is relaxed from 10~keV to about 2.7~keV
lower~\cite{Kusenko:2006rh}, as shown in Fig.~\ref{fig:one}.

Sterile neutrinos are stable on cosmological time scales, but they do decay
into the lighter neutrinos and the x-ray photons.  This two-body decay can be
used to discover the sterile dark matter.  The current x-ray
limits~\cite{x-rays} are shown in Fig.~\ref{fig:one} in two ways: (i) assuming
that the sterile neutrinos account for the entire dark matter (dashed line,
$\Omega_s\approx 0.2$), and (ii) assuming that the sterile neutrinos are only
produced via neutrino oscillations\cite{dw,Fuller}, in which case they may not
be the dominant dark-matter component. The latter scenario provides a
model-independent limit in any cosmology, except for the low-reheat
scenarios~\cite{low-reheat}. Regardless of how the sterile neutrinos are
produced, their decay width is related to the mixing angle, and to the fraction
of matter they make up. 

The same x-rays from sterile neutrinos can play an important role during the
dark ages in the early universe.  Although these x-rays alone are not
sufficient to reionize the universe~\cite{Mapelli:2005hq}, they can catalyze
the production of molecular hydrogen and speed up the star
formation~\cite{reion,reion1}, which, in turn, causes the reionization.

Molecular hydrogen is a very important
cooling agent necessary for the collapse of primordial gas clouds that give
birth to the first stars.  The fraction of molecular hydrogen must exceed a
certain minimal value for the star formation to begin~\cite{Tegmark:1996yt}. 
The reaction H+H$\rightarrow$H$_2 +\gamma$ is very slow in comparison with the
reaction 
\begin{eqnarray}
{\rm H}^{+}+{\rm H} & \rightarrow & {\rm H}_2^++ \gamma, \\
{\rm H}_2^{+}+{\rm H} & \rightarrow & {\rm H}_2+{\rm H}^+, 
\end{eqnarray}
which is possible if the hydrogen is ionized.  Therefore, the ionization
fraction $x_e$ determines the rate of molecular hydrogen production.  If dark
matter is made up of sterile neutrinos, their decays give out a sufficient flux
of photons to increase the ionization fraction by as much as two orders of
magnitude~\cite{reion,reion1,mapelli}.  This has a dramatic effect on the
production of molecular hydrogen and the subsequent star formation. 

\begin{figure*}[htb]
\includegraphics[width=36pc]{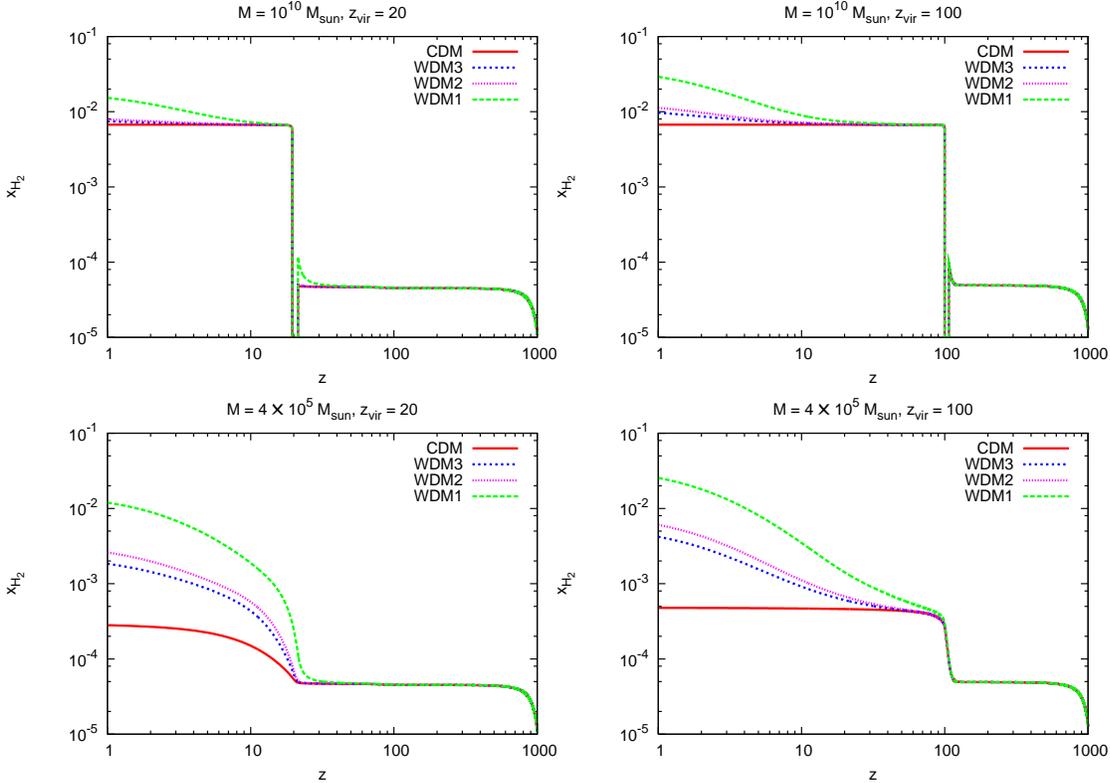}
\caption{Evolution of the molecular hydrogen fraction with
redshift~\cite{reion1} for three different models: $m_{s}
=25$~keV and $\sin^2 \theta=3 \times 10^{-12}$ (WDM1), $m_{s} =15$~keV and
$\sin^2 \theta=3 \times 10^{-12}$ (WDM2), $m_{s} =3.3$~keV and $\sin^2
\theta=3 \times 10^{-9}$ (WDM3). $M$ is the cloud mass and $z_{vir}$ is the 
redshift of virialization. 
}
\label{fig:H2}
\end{figure*}

The x-ray photons from the dark matter decays affect the collapsing gas halos
in two ways.  First, they catalyze the formation of molecular hydrogen, which
facilitates cooling.  Second, they heat up the gas clouds, which might
counteract cooling~\cite{reion1,mapelli}.  The interplay of these two effects
was studied in Ref.~\cite{reion1} for three points in the allowed parameter
space, WDM1-3, as shown in Fig.~\ref{fig:one}.  The results for the molecular
hydrogen fraction are shown in Fig.~\ref{fig:H2}.   In all three cases the the
main effect of sterile neutrinos was to facilitate the collapse of the clouds
and to speed up the star formation.

This work was supported in part by the DOE grant DE-FG03-91ER40662 and by the
NASA ATP grants NAG~5-10842 and NAG~5-13399.

\end{document}